\documentclass[10pt,conference]{IEEEtran}

\newtheorem{proposition}{Proposition}
\newtheorem{corollary}{Corollary}
\newtheorem{definition}{Definition}
\newtheorem{lemma}{Lemma}

\newcommand{\df}{\stackrel{\mbox{\scriptsize def}}{=}}
\newcommand{\ELS}{ELS}
\newcommand{\rk}{\mathrm{rk}}

\usepackage{amsmath, amssymb,cite}
\usepackage[dvips]{graphicx}

\begin{document}

\title{Decoder Error Probability of MRD Codes}
\author{\authorblockN{Maximilien Gadouleau}
\authorblockA{Department of Electrical and Computer Engineering\\
Lehigh University\\
Bethlehem, PA 18015 USA \\ E-mail: magc@lehigh.edu} \and
\authorblockN{Zhiyuan Yan}
\authorblockA{Department of Electrical and Computer Engineering\\
Lehigh University\\
Bethlehem, PA 18015 USA\\ E-mail: yan@lehigh.edu}
} \maketitle

\begin{abstract}
In this paper, we first introduce the concept of \emph{elementary
linear subspace}, which has similar properties to those of a set of
coordinates. Using this new concept, we derive properties of maximum
rank distance (MRD) codes that parallel those of maximum distance
separable (MDS) codes. Using these properties, we show that the
decoder error probability of MRD codes with error correction
capability $t$ \emph{decreases exponentially} with $t^2$ based on
the assumption that all errors with the same rank are equally
likely. We argue that the channel based on this assumption is an
approximation of a channel corrupted by crisscross errors.
\end{abstract}
\IEEEpeerreviewmaketitle

\section{Introduction}\label{sec:intro}
The Hamming metric has often been considered the most relevant
metric for error-control codes so far. Recently, the rank metric
\cite{gabidulin_pit0185} has attracted some attention due to its
relevance to space-time coding \cite{tarokh_98} and storage
applications \cite{roth_it91}. In \cite{lusina_it03}, space-time
block codes with good rank properties have been proposed. Rank
metric codes are used to correct crisscross errors that can be found
in memory chip arrays and magnetic tapes \cite{roth_it91}. Rank
metric codes have been used in public-key cryptosystems as well
\cite{gabidulin_lncs91}.

In \cite{gabidulin_pit0185}, a Singleton bound on the minimum rank
distance of rank metric codes was established, and codes that attain
this bound were called maximum rank distance (MRD) codes. An
explicit construction of MRD codes (these codes are referred to as
Gabidulin codes) was also given in \cite{gabidulin_pit0185}, and
this construction was extended in \cite{kshevetskiy_isit05}. Also, a
decoding algorithm that parallels the extended Euclidean algorithm
(EEA) was proposed for MRD codes.

In this paper, we investigate the performance of MRD codes when used
to protect data from additive errors based on two assumptions.
First, we assume all errors with the same rank are equally likely.
We argue that the channel based on this assumption is an
approximation of a channel corrupted by crisscross errors (see
Section~\ref{sec:main_results} for details). Second, we assume that
a bounded rank distance decoder is used, with error correction
capability $t$. If the error has rank no more than $t$, the decoder
gives the correct codeword. When the error has rank greater than
$t$, the output of the decoder is either a decoding failure or a
wrong codeword, which corresponds to a decoder error. Note that the
decoder error probability of maximum distance separable (MDS) codes
was investigated in \cite{mceliece_it86}, where all errors with the
same Hamming weight were assumed to be equiprobable. The main
contributions of this paper are:
\begin{itemize}
\item We introduce the concept of \emph{elementary
linear subspace} (ELS). The properties of an ELS are similar to
those of a set of coordinates.

\item Using elementary linear subspaces, we derive useful properties of MRD
codes. In particular, we prove the combinatorial property of MRD
codes, derive a bound on the rank distribution of these codes, and
show that the restriction of an MRD code on an ELS is also an MRD
code. These properties parallel those of MDS codes.

\item Using the properties of MRD codes, we derive a bound on the decoder error probability of
MRD codes that \emph{decreases exponentially} with $t^2$. Our
simulation results are consistent with our bound.
\end{itemize}

The rest of the paper is organized as follows.
Section~\ref{sec:preliminaries} gives a brief review of the rank
metric, Singleton bound, and MRD codes. In Section~\ref{sec:lemmas},
we first introduce the concept of elementary linear subspace and
study its properties, and then obtain some important properties of
MRD codes. Section~\ref{sec:main_results} derives the bound on the
decoder error probability of MRD codes when all errors with the same
rank are equiprobable. Finally, in Section~\ref{sec:simulations},
our bound on the decoder error probability is confirmed by
simulation results.

\section{Preliminaries}\label{sec:preliminaries}
\subsection{Rank metric}
Consider an $n$-dimensional vector ${\bf x} = (x_0, x_1,\ldots,
x_{n-1})$ in $\mathrm{GF}(q^m)^n$. Assume $\alpha_0, \alpha_1,
\ldots, \alpha_{m-1}$ is a basis set of GF$(q^m)$ over GF$(q)$, then
for $j=0, 1, \ldots, n-1$, $x_j$ can be written as $x_j =
\sum_{i=0}^{m-1} x_{i, j}\alpha_i$, where $x_{i, j} \in
\mbox{GF}(q)$ for $i=0, 1, \ldots, m-1$. Hence, $x_j$ can be
expanded to an $m$-dimensional column vector $(x_{0, j}, x_{1,
j},\ldots, x_{m-1, j})^T$ with respect to the basis set $\alpha_0,
\alpha_1, \ldots, \alpha_{m-1}$. Let ${\bf X}$ be the $m\times n$
matrix obtained by expanding all the coordinates of ${\bf x}$. That
is,
\begin{displaymath}
    {\bf X} = \left(
    \begin{array}{cccc}
        x_{0, 0} & x_{0, 1} & \ldots & x_{0, n-1}\\
        x_{1, 0} & x_{1, 1} & \ldots & x_{1, n-1}\\
        \vdots & \vdots & \ddots & \vdots\\
        x_{m-1, 0} & x_{m-1, 1} & \ldots & x_{m-1, n-1}
    \end{array}
    \right),
\end{displaymath}
where $x_j = \sum_{i=0}^{m-1} x_{i, j}\alpha_i$.  The \emph{rank
norm} of the vector ${\bf x}$ (over GF$(q)$), denoted as $\rk({\bf
x}|\mbox{GF}(q))$, is defined to be the rank of the matrix ${\bf X}$
over GF$(q)$, i.e., $\rk({\bf x}|\mbox{GF}(q)) \df
\mathrm{rank}({\bf X})$ \cite{gabidulin_pit0185}. The rank norm of
${\bf x}$ can also be viewed as the smallest number of rank 1
matrices ${\bf B}_i$ such that ${\bf X} = \sum_i {\bf B}_i$
\cite{gabidulin_itct03}. All the ranks are over the base field
GF$(q)$ unless otherwise specified in this paper. To simplify
notations, we denote the rank norm of ${\bf x}$ as $\rk({\bf x})$
henceforth. Accordingly, $\forall \, {\bf x}, {\bf y}\in
\mathrm{GF}(q^m)^n$, $d({\bf x},{\bf y})\df \rk({\bf x} - {\bf y})$
is shown to be a metric over GF$(q^m)^n$, referred to as \emph{the
rank metric} henceforth \cite{gabidulin_pit0185}. Hence, the {\em
minimum rank distance} $d$ of a code of length $n$ is simply the
minimum rank distance over all possible pairs of distinct codewords.
A code with a minimum rank distance $d$ can correct errors with rank
up to $t = \left\lfloor (d-1)/2 \right\rfloor$.

\subsection{The Singleton bound and MRD codes}
The minimum rank distance of a code can be specifically bounded.
First, the minimum rank distance $d$ of a code over
$\mathrm{GF}(q^m)$ is obviously bounded by $m$. Codes that satisfy
$d = m$ are studied in \cite{manoj_report1002}. Also, it can be
shown that $d \leq d_{\mbox{\tiny H}}$ \cite{gabidulin_pit0185},
where $d_{\mbox{\tiny H}}$ is the minimum Hamming distance of the
same code. Due to the Singleton bound on the minimum Hamming
distance of block codes, the minimum rank distance of an $(n,M)$
block code over $\mathrm{GF}(q^m)$ thus satisfies
\begin{equation}\label{eq:singleton1}
    d\leq n-\log_{q^m}M+1.
\end{equation}
In this paper, we refer to the bound in (\ref{eq:singleton1}) as the
Singleton bound for rank metric codes and codes that attain the
bound as MRD codes. Note that (\ref{eq:singleton1}) implies that the
cardinality of any MRD code is a power of $q^m$.

\section{Properties of MRD codes}\label{sec:lemmas}
In order to derive the bound on decoder error probability, we need
some important properties of MRD codes, which will be established
next using the concept of elementary linear subspace.

\subsection{Elementary linear subspaces}\label{sec:ELS}
Many properties of MDS codes are established by studying sets of
coordinates. These sets of coordinates may be viewed as linear
subspaces which have a basis of vectors with Hamming weight $1$. We
first define elementary linear subspaces, which are the counterparts
of sets of coordinates.

\begin{definition}[Elementary linear subspace (\ELS{})]
A linear subspace $V$ of $\mathrm{GF}(q^m)^n$ is said to be
elementary if it has a basis $\mathcal{B}$ consisting of row vectors
in $\mathrm{GF}(q)^n$. $\mathcal{B}$ is called an elementary basis
of $V$.
\end{definition}

We remark that $V$ is an ELS if and only if it has a basis
consisting of vectors of rank $1$. We also remark that not all
linear subspaces are elementary. For example, the span of a vector
of rank $u>1$ has dimension $1$, but requires $u$ vectors of rank
$1$ to span it. Next, we show that the properties of elementary
linear subspaces are similar to those of sets of coordinates.
\begin{definition}[Rank of a linear subspace]
The rank of a linear subspace $L$ of $\mathrm{GF}(q^m)^n$ is defined
to be the maximum rank among the vectors in $L$
$$
    \rk(L) \df \max_{{\bf x} \in L}\{\rk({\bf x})\}.
$$
\end{definition}

\begin{proposition}\label{prop:rk=dim}
If $V$ is an \ELS{} of $\mathrm{GF}(q^m)^n$ with $\mathrm{dim}(V)
\leq m$, then $\rk(V) = \mathrm{dim}(V).$
\end{proposition}

\begin{proof}
Let us first denote $\mathrm{dim}(V)$ as $v$. Any vector ${\bf x}
\in V$ can be expressed as the sum of at most $v$ vectors of rank
$1$, hence its rank is upper bounded by $v$. Thus, $\rk(V) \leq v$
and it suffices to find a vector in $V$ with rank equal to $v$. Let
$\mathcal{B} = \{ {\bf b}_j \}_{j=0}^{v-1}$ be an elementary basis
of $V$, and consider ${\bf y} = \sum_{i=0}^{v-1} \alpha_i {\bf
b}_i$, where $\{\alpha_i \}_{i=0}^{v-1}$ all belong to a basis set
of $\mathrm{GF}(q^m)$ over GF$(q)$. If we expand the coordinates of
${\bf y}$ with respect to the basis $\{ \alpha_i \}_{i=0}^{v-1}$, we
get
$$
    {\bf Y} = \left({\bf b}_0^T,\ldots,{\bf b}_{v-1}^T,{\bf 0}^T,\ldots,
    {\bf 0}^T\right)^T.
$$
Since the row vectors ${\bf b}_0, {\bf b}_1, \cdots, {\bf b}_{v-1}$
are linearly independent over GF$(q)$, ${\bf Y}$ has rank $v$ and
$\rk({\bf y}) = v$.
\end{proof}
\begin{lemma}\label{lemma:x_in_A}
Any vector ${\bf x} \in \mathrm{GF}(q^m)^n$ with rank $u$ belongs to
an \ELS{} $A$ of dimension $u$. Also, ${\bf x}$ is not contained in
any \ELS{} of dimension less than $u$.
\end{lemma}

\begin{proof}
The vector ${\bf x}$ can be expressed as a linear combination ${\bf
x} = \sum_{i=0}^{u-1} x_i{\bf a}_i$, where $x_i \in
\mathrm{GF}(q^m)$ and ${\bf a}_i \in \mathrm{GF}(q)^n$ for $0 \leq i
\leq u-1$. Let $A$ be the ELS of $\mathrm{GF}(q^m)^n$ spanned by
${\bf a}_i$'s, then $\mathrm{dim}(A)=u$ and ${\bf x} \in A$. Also,
suppose ${\bf x} \in B$, where $B$ is an \ELS{} with dimension $t <
u$. Then by Proposition~\ref{prop:rk=dim}, $\rk({\bf x}) \leq t <
u$, which is a contradiction.
\end{proof}

\begin{lemma}\label{lemma:f}
Let $\mathcal{V}$ be the set of all \ELS{}'s of $\mathrm{GF}(q^m)^n$
with dimension $v$ and $\mathcal{V}'$ be the set of all linear
subspaces of $\mathrm{GF}(q)^n$ with dimension $v$. Then there
exists a bijection between $\mathcal{V}$ and $\mathcal{V}'$.
\end{lemma}
\begin{proof}
Let $V$ be an \ELS{} of $\mathrm{GF}(q^m)^n$ with $\mathcal{B}$ as
an elementary basis. For any positive integer $a$, define
$$
    \left<\mathcal{B}\right>_{q^a} \df \left\{ \left. \sum_{i=0}^{v-1} x_i
    {\bf b}_i \,\right|\, x_i \in \mathrm{GF}(q^a), {\bf b}_i \in
    \mathcal{B} \right\}
$$
and a mapping $f:\mathcal{V}\rightarrow \mathcal{V}'$ given by $V =
\left<\mathcal{B}\right>_{q^m} \mapsto  V' =
    \left<\mathcal{B}\right>_{q}$.
This mapping takes $V$ to a unique $V'$, which is the set of all
linear combinations over $\mathrm{GF}(q)$ of the vectors in
$\mathcal{B}$. Note that $\mathrm{dim}(V') = \mathrm{dim}(V)$. $f$
is injective because if $V$ and $W$ are distinct, then $V'$ and $W'$
are also distinct. $f$ is also surjective since for all $V' \in
\mathcal{V}'$, there exists $V \in \mathcal{V}$ such that $f(V) =
V'$.
\end{proof}
For $0 \leq v \leq n$, let ${n \brack v}$ be the number of linear
subspaces of $\mathrm{GF}(q)^n$ with dimension $v$ \cite{andrews}.
For $v=0$, ${n \brack 0} = 1$, and for $1 \leq v \leq n$: ${n \brack
v} = \prod_{i=0}^{v-1} \frac{q^n-q^i}{q^v-q^i}.$ Lemma~\ref{lemma:f}
implies that
\begin{corollary}\label{cor:number_ELS}
There are ${n \brack v}$ \ELS{}'s in $\mathrm{GF}(q^m)^n$ with
dimension $v$.
\end{corollary}

\begin{proposition}\label{prop:complementary_ELS}
Let $V$ be an \ELS{} of $\mathrm{GF}(q^m)^n$ with dimension $v$,
then there exists an \ELS{} $W$ complementary to $V$, i.e., $V
\oplus W = \mathrm{GF}(q^m)^n$.
\end{proposition}

\begin{proof}
Let $V' = f(V)$, and let $\mathcal{B}$ a basis of $V'$, then there
exists $W'$, with basis $\bar{\mathcal B}$, such that $V' \oplus W'
= \mathrm{GF}(q)^n$. Denote $W = f^{-1}(W')$. We now want to show
that $W \oplus V = \mathrm{GF}(q^m)^n$. First, $\mathrm{dim}(W)$ is
$n-v$, hence we only need to show that $V \cap W = \{{\bf 0}\}$. Let
${\bf y}\neq {\bf 0} \in V \cap W$, then there is a non-trivial
linear relationship over $\mathrm{GF}(q^m)$ among the elements of
$\mathcal{B}$ and those of $\bar{\mathcal B}$. This may be expressed
as
\begin{equation}\sum_{{\bf b}_i\in \mathcal{B} \cup \bar{\mathcal B}}y_i {\bf b}_i = {\bf
0}.\label{eq:complementary_ELS}\end{equation}
Applying the trace
function to each coordinate on both sides of
(\ref{eq:complementary_ELS}), we obtain $\sum_{{\bf b}_i\in
\mathcal{B} \cup \bar{\mathcal B}} \mathrm{Tr}(y_i) {\bf b}_i = {\bf
0}$,  which implies a linear dependence over $\mathrm{GF}(q)$ of the
vectors in $\mathcal{B}$ and $\bar{\mathcal B}$. This contradicts
the fact that $\mathcal{B} \cup \bar{\mathcal B}$ is a basis.
Therefore, $W \oplus V = \mathrm{GF}(q^m)^n$.
\end{proof}

\begin{definition}[Restriction of a vector]
Let $L$ be a linear subspace of $\mathrm{GF}(q^m)^n$ and let ${\bar
L}$ be complementary to $L$, i.e., $L \oplus \bar{L} =
\mathrm{GF}(q^m)^n$. Any vector ${\bf x} \in \mathrm{GF}(q^m)^n$ can
then be represented as ${\bf x} = {\bf x}_L \oplus {\bf
x}_{\bar{L}}$, where ${\bf x}_L \in L$ and ${\bf x}_{\bar{L}} \in
{\bar L}$. We will call ${\bf x}_L$ and ${\bf x}_{\bar L}$ the
restrictions of ${\bf x}$ on $L$ and ${\bar L}$, respectively.
\end{definition}
Note that for any given linear subspace $L$, its complementary
linear subspace $\bar{L}$ is not unique. Furthermore, the
restriction of ${\bf x}$ on $L$, ${\bf x}_L $, depends on not only
$L$ but also ${\bar L}$. Thus, ${\bf x}_L $ is well defined only
when both $L$ and ${\bar L}$ are given. All the restrictions in this
paper are with respect to a fixed pair of linear subspaces
complementary to each other.
\begin{definition}
Let ${\bf x} \in \mathrm{GF}(q^m)^n$ and $V$ be an \ELS{}. If there
exists an \ELS{} ${\bar V}$ complementary to $V$ such that ${\bf x}
= {\bf x}_{\bar V}\oplus {\bf 0}$, we say that ${\bf x}$ vanishes on
$V$.
\end{definition}

\begin{lemma}\label{lemma:vanish}
Given a vector ${\bf x} \in \mathrm{GF}(q^m)^n$ of rank $u$, there
exists an \ELS{} with dimension $n-u$ on which ${\bf x}$ vanishes.
Also, ${\bf x}$ does not vanish on any \ELS{} with dimension greater
than $n-u$.
\end{lemma}

\begin{proof}
By Lemma~\ref{lemma:x_in_A}, ${\bf x} \in A$, where $A$ is an \ELS{}
with dimension $u$. Let ${\bar A}$ be an \ELS{} with dimension $n-u$
that is complementary to $A$. Thus, ${\bf x}$ may be expressed as
${\bf x} = {\bf x}_A \oplus {\bf x}_{\bar A}={\bf x}_A \oplus {\bf
0}$. That is, ${\bf x}$ vanishes on ${\bar A}$. Also, suppose ${\bf
x}$ vanishes on an \ELS{} $B$ with dimension greater than $n-u$.
Then there exists an \ELS{} ${\bar B}$ with dimension $< u$ such
that ${\bf x} \in {\bar B}$, which contradicts
Lemma~\ref{lemma:x_in_A}.
\end{proof}

\subsection{Properties of MRD codes}\label{sec:properties_MRD}
We now derive some useful properties for MRD codes, which will be
used in our derivation of the decoder error probability. In this
subsection, let $C$ be an MRD code over $\mathrm{GF}(q^m)$ with
length $n$ ($n \leq m$) and cardinality $q^{mk}$. Note that $C$ may
be linear or nonlinear. First, we derive the basic combinatorial
property of MRD codes.
\begin{lemma}[Basic combinatorial property]
\label{lemma:combinatorial_prop} Let $K$ be an \ELS{} of
$\mathrm{GF}(q^m)^n$ with dimension $k$, and fix ${\bar K}$, an ELS
complementary to $K$. Then, for any vector ${\bf k} \in K$, there
exists a unique codeword ${\bf c} \in C$ such that its restriction
on $K$ satisfies ${\bf c}_K = {\bf k}$.
\end{lemma}
\begin{proof}
Suppose there exist ${\bf c}\ne {\bf d} \in C$ such that ${\bf c}_K
= {\bf d}_K$. Their difference ${\bf c} - {\bf d}$ is in ${\bar K}$,
and hence has rank at most $n-k$ by Proposition~\ref{prop:rk=dim},
which contradicts the fact that $C$ is MRD. Then all the codewords
lead to different restrictions on $K$. Also, $|C| =|K| = q^{mk}$,
thus for any ${\bf k}$, there exists a unique ${\bf c}$ such that
${\bf c}_K = {\bf k}$.
\end{proof}

This property allows us to obtain a bound on the rank distribution
of MRD codes.

\begin{lemma}[Bound on the rank distribution]\label{lemma:bound_Au}
Let $A_u$ be the number of codewords of $C$ with rank $u$. Then, for
the redundancy $r = n-k$ and $u \geq d$,
\begin{equation}\label{eq:bound_Au}
    A_u \leq {n \brack u}(q^m-1)^{u-r}.
\end{equation}
\end{lemma}
\begin{proof}
From Lemma~\ref{lemma:vanish}, any codeword ${\bf c}$ with rank $u
\geq d$ vanishes on an \ELS{} with dimension $v=n-u$. Thus
(\ref{eq:bound_Au}) can be established by first determining the
number of codewords vanishing on a given \ELS{}, and then
multiplying by the number of such \ELS{}'s, ${n \brack u}$. Let $V$
be an arbitrary \ELS{} with dimension $v$. First, since $v \leq
k-1$, $V$ is properly contained in an \ELS{} $K$ with dimension $k$.
According to the combinatorial property, ${\bf c}$ is completely
determined by ${\bf c}_K$. Hence, if we specify that a codeword
vanishes on $V$, we may specify $k-v$ other nonzero components
arbitrarily. There are at most $(q^m-1)^{k-v} = (q^m-1)^{u-r}$ ways
to do so, implying that there are at most $(q^m-1)^{u-r}$ vectors
that vanish on $V$.
\end{proof}

Note that the exact formula for the rank distribution of linear MRD
codes was derived in \cite{gabidulin_pit0185}. However, the bound in
(\ref{eq:bound_Au}) is more convenient for the present application.

It is well known that a punctured MDS code is an MDS code
\cite{blahut_83}. We will show that the restriction of an MRD code
to an \ELS{} is also MRD. Let $V$ be an \ELS{} with dimension $v
\geq k$, an elementary basis $\left\{{\bf b}_0, {\bf b}_1, \cdots,
{\bf b}_{v-1}\right\}$, and a complementary ELS $\bar{V}$. For any
codeword ${\bf c}$, suppose ${\bf c}_V=\sum_{i=0}^{v-1}a_i{\bf
b}_i$, where $a_i\in \mbox{GF}(q^m)$. Let us define a mapping $r:
\mbox{GF}(q^m)^n\rightarrow \mbox{GF}(q^m)^v$ given by ${\bf
c}\mapsto r({\bf c})=\left(a_0, \cdots, a_{v-1}\right)$. Then $C_V =
\{ r({\bf c}) | {\bf c} \in C\}$ is called the restriction of $C$ on
$V$.
\begin{lemma}[Restriction of an MRD
code]\label{lemma:restriction_MRD} For an \ELS{} $V$ with dimension
$v \geq k$, $C_V$ is an MRD code.
\end{lemma}

\begin{proof}
Clearly, $C_V$ is a code over $\mathrm{GF}(q^m)$ with length $v$ ($v
\leq m$) and cardinality $q^{mk}$. Assume ${\bf c}\ne {\bf d} \in
C$, and consider ${\bf x} = {\bf c} - {\bf d}$. Then we have
$\rk(r({\bf c})-r({\bf d})) =\rk({\bf x}_V) \geq \rk({\bf x}) -
\rk({\bf x}_{\bar{V}})\geq n-k+1 - (n-v) = v-k+1$. The Singleton
bound on $C_V$ completes the proof.
\end{proof}

\section{Decoder error probability of MRD codes in case of crisscross errors}
\label{sec:main_results} Let $C$ be a linear $(n, k)$ MRD code over
$\mathrm{GF}(q^m)$ ($n\leq m$) with minimum rank distance $d=n-k+1$
and error correction capability $t = \left\lfloor (d-1)/2
\right\rfloor$. We assume that $C$ is used to protect data from
additive errors with rank $u$. That is, the received word
corresponding to a codeword ${\bf c}$ of $C$ is ${\bf c}+{\bf e}$.
We argue that the additive error with rank $u$ is an approximation
of crisscross errors. Let us assume $q=2$ and a codeword can be
represented by an $m \times n$ array of bits. Suppose some of the
bits are recorded erroneously, and the error patterns are such that
all corrupted bits are confined to a number of rows or columns or
both. Such an error model, referred to as crisscross errors, occurs
in memory chip arrays or magnetic tapes \cite{roth_it91}. Suppose
the errors are confined to a row (or column), then such an error
pattern can be viewed as the addition of an error array which has
non-zero coordinates on only one row (or column) and hence has rank
$1$. We may reasonably assume each row is corrupted equally likely
and so is each column. Thus, all the errors that are restricted to
$u>1$ rows (or columns) are equally likely. Finally, if we assume
the probability of a corrupted row is the same as that of a
corrupted column, then all crisscross errors with weight $u$
\cite{roth_it91} are equally likely. The weight $u$ of the
crisscross error is no less than the rank of the error
\cite{roth_it91}. However, in many cases the weight $u$ equals the
rank. Hence, assuming all errors with the same rank are equiprobable
is an approximation of crisscross errors.

A bounded distance decoder, which looks for a codeword within rank
distance $t$ of the received word, is used to correct the error.
Clearly, if ${\bf e}$ has rank no more than $t$, the decoder gives
the correct codeword. When the error has rank greater than $t$, the
output of the decoder is either a decoding failure or a decoder
error. We denote the probabilities of error and failure --- for
error correction capability $t$ and a given error rank $u$ --- as
$P_E(t; u)$ and $P_F(t; u)$ respectively. If $u \leq t$, then
$P_F(t; u)=P_E(t; u)=0$. When $u > t$, $P_E(t; u) + P_F(t; u) = 1$.
In particular, if $t < u < d-t$, then $P_E(t; u)=0$ and $P_F(t;
u)=1$; Thus we only need to investigate the case where $u \geq d-t$.

Since $C$ is linear and hence geometrically uniform, we assume
without loss of generality that the all-zero codeword is sent. Thus,
the received word can be any vector with rank $u$ with equal
probability. We call a vector decodable if it lies within rank
distance $t$ of some codeword. If $D_u$ denotes the number of
decodable vectors of rank $u$, then for $u \geq t+1$ we have
\begin{equation}\label{eq:PE_Du_Nu}
    P_E(t;u) = \frac{D_u}{N_u}=\frac{D_u}{{n
\brack u}A(m,u)},
\end{equation}
where $N_u$ denotes the number of vectors of rank $u$ and $A(m,u)
\df \prod_{i=0}^{u-1}(q^m-q^i)$. Hence the main challenge is to
derive upper bounds on $D_u$. We have to distinguish two cases: $u
\geq d$ and $u < d$. The approach we use to bound $D_u$ is similar
to that in \cite{mceliece_it86}.

\begin{proposition}\label{prop:Du_u_geq_d}
For $u \geq d$, then
\begin{equation}\label{eq:Du1}
    D_u \leq {n \brack u}(q^m-1)^{u-r}V_t,
\end{equation}
where $V_t=\sum_{i=0}^t N_i$ is the volume of a ball of rank radius
$t$.
\end{proposition}

\begin{proof}
Each decodable vector can be written uniquely as ${\bf c} + {\bf
e}$, where ${\bf c} \in C$ and $\rk({\bf e}) \leq t$. For a fixed
${\bf e}$, $C + {\bf e}$ is an MRD code, so it satisfies
Equation~(\ref{eq:bound_Au}). Therefore, the number of decodable
words of rank $u$ is at most ${n \brack u}(q^m-1)^{u-r}$ multiplied
by the number of possible error vectors, $V_t$.
\end{proof}

\begin{lemma}\label{lemma:choices_z}
Given ${\bf y} \in \mathrm{GF}(q^m)^v$ with rank $w$, there are at
most ${u \brack s-w} A(m,s-w) q^{w(u-s+w)}$ vectors ${\bf z} \in
\mathrm{GF}(q^m)^{n-v}$ such that ${\bf x}= ({\bf y}, {\bf z}) \in
\mathrm{GF}(q^m)^n$ has rank $s$.
\end{lemma}
\begin{proof}
The vector ${\bf x}$ has $s$ linearly independent coordinates. Since
$w$ of them are in ${\bf y}$, then $s-w$ of them are in ${\bf z}$.
Thus ${\bf z}$ has $s-w$ linearly independent coordinates which do
not belong to $\mathfrak{S}({\bf y})$. Without loss of generality,
assume those coordinates are on the first $s-w$ positions of ${\bf
z}$, and denote these coordinates as ${\bf z}'$. For $s-w+1 \leq i
\leq n-v$, $z_i$ is a linear combination of the coordinates of ${\bf
y}$ and the first $s-w$ coordinates of ${\bf z}$. Hence, we have
$z_i = a_i + b_i$, where $a_i \in \mathfrak{S}({\bf y})$ and $b_i
\in \mathfrak{S}({\bf z}')$. There are $q^{w(u-s+w)}$ choices for
the vector ${\bf a} = (0,\ldots,0,a_{s-w+1},\ldots,a_u)$. The vector
${\bf b}$ has rank $s-w$, so there are at most ${u \brack s-w}
A(m,s-w)$ choices for ${\bf b}$.
\end{proof}

\begin{proposition}\label{prop:Du_u<d}
For $d-t \leq u < d$, we have
\begin{eqnarray}\label{eq:Du_u<d}
    \nonumber
    D_u &\leq& {n \brack u} \sum_{w=d-u}^t {v \brack
    w}(q^m-1)^{w-r'}\cdots\\
    && \sum_{s=w}^t {u \brack s-w} A(m,s-w)q^{w(u-s+w)}.
\end{eqnarray}
\end{proposition}

\begin{proof}
Recall that a decodable vector of rank $u$ can be expressed as ${\bf
c} + {\bf e}$, where ${\bf c} \in C$ and $\rk({\bf e}) \leq t$. This
vector vanishes on an \ELS{} $\mathcal{V}$ with dimension $v = n-u$
by Lemma~\ref{lemma:vanish}. Fix $\bar{\mathcal{V}}$, an \ELS{}
complementary to $\mathcal{V}$. We have $w \df
\rk(r_\mathcal{V}({\bf c})) \leq t$. $C_\mathcal{V}$ is an MRD code
by Lemma~\ref{lemma:restriction_MRD}, hence $w \geq d-u$. By
Lemma~\ref{lemma:bound_Au}, and denoting $r'=r-u$, the number of
codewords of $C_\mathcal{V}$ with rank $w$ is at most ${v \brack
w}(q^m-1)^{w-r'}$. For each codeword ${\bf c}$ such that
$\rk(r_\mathcal{V}({\bf c})) = w$, we must count the number of error
vectors ${\bf e}$ such that $r_\mathcal{V}({\bf c}) +
r_\mathcal{V}({\bf e}) = {\bf 0}$. Suppose that ${\bf e}$ has rank
$s \geq w$, and denote ${\bf g} = r_\mathcal{V}({\bf e})$ and ${\bf
f} = r_{\bar{\mathcal{V}}}({\bf e})$. Note that ${\bf e}$ is
completely determined by ${\bf f}$. The vector $({\bf g},{\bf f})$
has rank $s$, hence by Lemma~\ref{lemma:choices_z}, there are at
most ${u \brack s-w} A(m,s-w)$ choices for the vector ${\bf f}$.

The total number $D_\mathcal{V}$ of decodable vectors vanishing on
$\mathcal{V}$ is then at most
\begin{eqnarray}\label{eq:DV_u<d}
\nonumber
    D_\mathcal{V} &\leq& \sum_{w=d-u}^t {v \brack
    w}(q^m-1)^{w-r'}\cdots\\
    &&\sum_{s=w}^t {u \brack s-w}A(m,s-w) q^{w(u-s+w)}.
\end{eqnarray}
The number of possible \ELS{}'s of dimension $v$ is ${n \brack v}$.
Multiplying the bound on $D_\mathcal{V}$ by ${n \brack v}$, the
number of possible \ELS{}'s of dimension $v$, we get the result.
\end{proof}

We can obtain a bound similar to~(\ref{eq:Du1}) which applies to the
case $d-t \leq u < d$.

\begin{corollary}\label{cor:bound_Du}
For $d- t\leq u < d$, then $D_u < \frac{q^2}{q^2-1}{n \brack
u}(q^m-1)^{u-r}V_t$.
\end{corollary}

\begin{proof}
We shall use Equation~(\ref{eq:DV_u<d}). We have
\begin{eqnarray}
    \nonumber
    D_\mathcal{V} &\leq& (q^m-1)^{u-r} \sum_{s=w}^t \sum_{w=d-u}^s
    {v \brack w} {u \brack s-w} \cdots \\
    && q^{w(u-s+w)} A(m,s-w)(q^m-1)^w\\
    \nonumber
    &<& (q^m-1)^{u-r}\sum_{s=d-u}^t q^{ms} \cdots \\
    && \sum_{w=d-u}^s {v \brack w} {u \brack s-w} q^{w(u-s+w)}.
\end{eqnarray}
Using the following combinatorial relation \cite{andrews}:
$\sum_{w=d-u}^s {v \brack w} {u \brack s-w} q^{w(u-s+w)} = {v+u
\brack s}$, we obtain $D_\mathcal{V} < (q^m-1)^{u-r}\sum_{s=d-u}^t
q^{ms} {n \brack s}$. It can be shown that $q^{ms} \leq
\frac{q^2}{q^2-1}A(m,s)$. Using this result, we find that
$D_\mathcal{V} < \frac{q^2}{q^2-1} (q^m-1)^{u-r}\sum_{s=d-u}^t
A(m,s) {n \brack s} < \frac{q^2}{q^2-1} (q^m-1)^{u-r} V_t$.
\end{proof}

We can eventually derive a bound on the decoder error probability.
\begin{proposition}\label{prop:bound_PE1}
For $d-t \leq u < d$, the decoder error probability satisfies
\begin{equation}\label{eq:bound_PE11}
    P_E(t;u) < \frac{q^2}{q^2-1} \frac{(q^m-1)^{u-r}}{A(m,u)}V_t.
\end{equation}
For $u \geq d$, the decoder error probability satisfies
\begin{equation}\label{eq:bound_PE12}
    P_E(t;u) < \frac{(q^m-1)^{u-r}}{A(m,u)}V_t.
\end{equation}
\end{proposition}
\begin{proof}
Directly from Proposition~\ref{prop:Du_u_geq_d} and
Corollary~\ref{cor:bound_Du}.
\end{proof}

Before deriving an upper bound for $P_E(t;u)$, we need to establish
two lemmas.
\begin{lemma}\label{lemma:bound_Amu}
For $0 \leq u \leq m$, $A(m,u) \geq q^{mu-\sigma(q)}$, where
$\sigma(q) = \frac{1}{\ln(q)} \sum_{k=1}^\infty \frac{1}{k(q^k-1)}$
is a decreasing function of $q$ with $\sigma(2) \approx 1.7919$.
\end{lemma}
\begin{proof} We have $A(m,u) = q^{mu-M_u}$,
where $M_u = -\sum_{j=m-u+1}^m \log_q(1-q^{-j})$. $M_u$ is an
increasing function of $u$, with maximum equal to $M_m =
\frac{1}{\ln(q)}\sum_{k=1}^{\infty}\frac{1-q^{-mk}}{k(q^k-1)} \leq
\sigma(q)$.
\end{proof}

\begin{lemma}\label{lemma:bound_Vt}
For $m \geq 1$ and $t \leq m/2$, $V_t \leq q^{t(n+m-t)+\sigma(q)}$.
\end{lemma}
\begin{proof}First, we need to prove the following claim.

Claim: For $m \geq 1$ and $i \leq t \leq m/2$, we have ${m \brack
t-i} q^{-i(t-i)} \geq 1$.

An exhaustive search proves the result for $m < 4$. We shall assume
that $m \geq 4$ herein. The case $i=t$ being trivial, we hence
assume that $i < t$. Using Lemma~\ref{lemma:bound_Amu}, we find that
${m \brack t-i} \geq q^{(t-i)(m-t+i)-\sigma(q)}$, hence ${m \brack
t-i} q^{-i(t-i)}\geq q^{(t-i)(m-t)-\sigma(q)} \geq q^{m/2-\sigma(q)}
\geq 1$.

The claim implies $A(m,i) \leq q^{mi} \leq {m \brack
t-i}q^{i(m-t+i)}$. Since $V_t = \sum_{i=0}^t {n \brack i}A(m,i)$,
the bound on $A(m,i)$ allows us to derive the bound on $V_t$.
\end{proof}

The result in Proposition~\ref{prop:bound_PE1} may be weakened in
order to find a bound on the decoder error probability which only
depends on $t$.
\begin{proposition}\label{prop:bound_PE2}
For $u \geq d-t$, the decoder error probability satisfies
\begin{equation}\label{eq:bound_PE3}
    P_E(t;u) < q^{-t^2+2\sigma(q)}.
\end{equation}
\end{proposition}
\begin{proof}
First suppose that $u \geq d$. From
Proposition~\ref{prop:bound_PE1}, we have for $u \geq d-t$: $
P_E(t;u) < \frac{(q^m-1)^{u-r}}{A(m,u)}V_t.$  The bounds in
Lemmas~\ref{lemma:bound_Amu} and \ref{lemma:bound_Vt} lead to
$P_E(t;u) < q^{-mr+t(m+n-t)+2\sigma(q)}$. Since $n \leq m$ and $2t
\leq r$, it follows that $P_E(t;u) < q^{-t^2 + 2\sigma(q)}$. For
$d-t \leq u < d$, we find that $P_E(t;u) < \frac{q^2(q-1)}{q(q^2-1)}
q^{-mr+t(m+n-t)+2\sigma(q)} < q^{-mr+t(m+n-t)+2\sigma(q)}$. Using
the same reasoning as above, we find the same conclusion.
\end{proof}

\section{Simulation Results}\label{sec:simulations}
In this section, we use numerical simulations to verify our bound
given in Proposition~\ref{prop:bound_PE2}. In our simulations, we
used a special family of MRD codes called Gabidulin codes
\cite{gabidulin_pit0185} with the following parameters: $q=2$,
$m=n=16$, and $d=2t+1=n-k+1$. The simulations were based on the
following process: first a random message word in
$\mathrm{GF}(q^m)^k$ is encoded using the generator matrix of the
Gabidulin code, then a random error vector with rank $u>t$ is added
to the codeword, and finally the EEA \cite{gabidulin_pit0185} is
used to decode the received word. Since $u>t$, the decoding results
in either a failure or an error. Similarly to the decoding of
Reed-Solomon codes, decoder failure is declared based on the output
of the EEA. Different values for $t$ and $u$ are used in our
simulations to verify our bound. Each value of the decoder error
probability is computed after at least $15$ occurrences of decoder
errors to ensure reliability of simulation results.

\begin{figure}[htp]
\begin{center}
\includegraphics[scale=0.65]{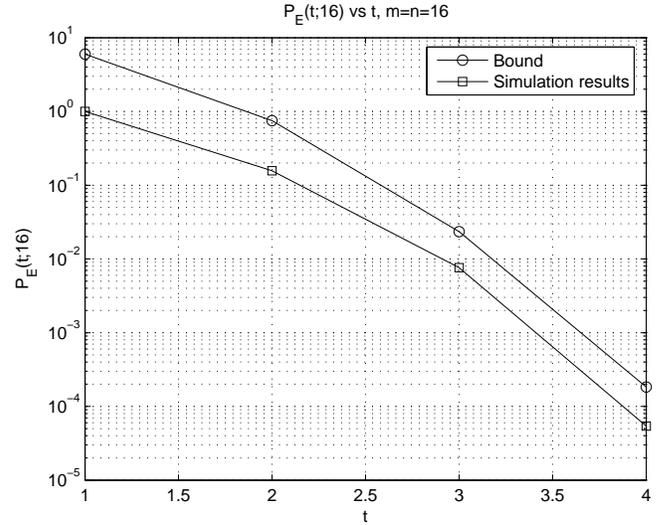}
\end{center}
\caption{Decoder error probability of an MRD code as a function of
$t$, with $q=2$, $m=n=16$, and $u=16$.}\label{fig:PE_MRDt}
\end{figure}

\begin{figure}[htp]
\begin{center}
\includegraphics[scale=0.65]{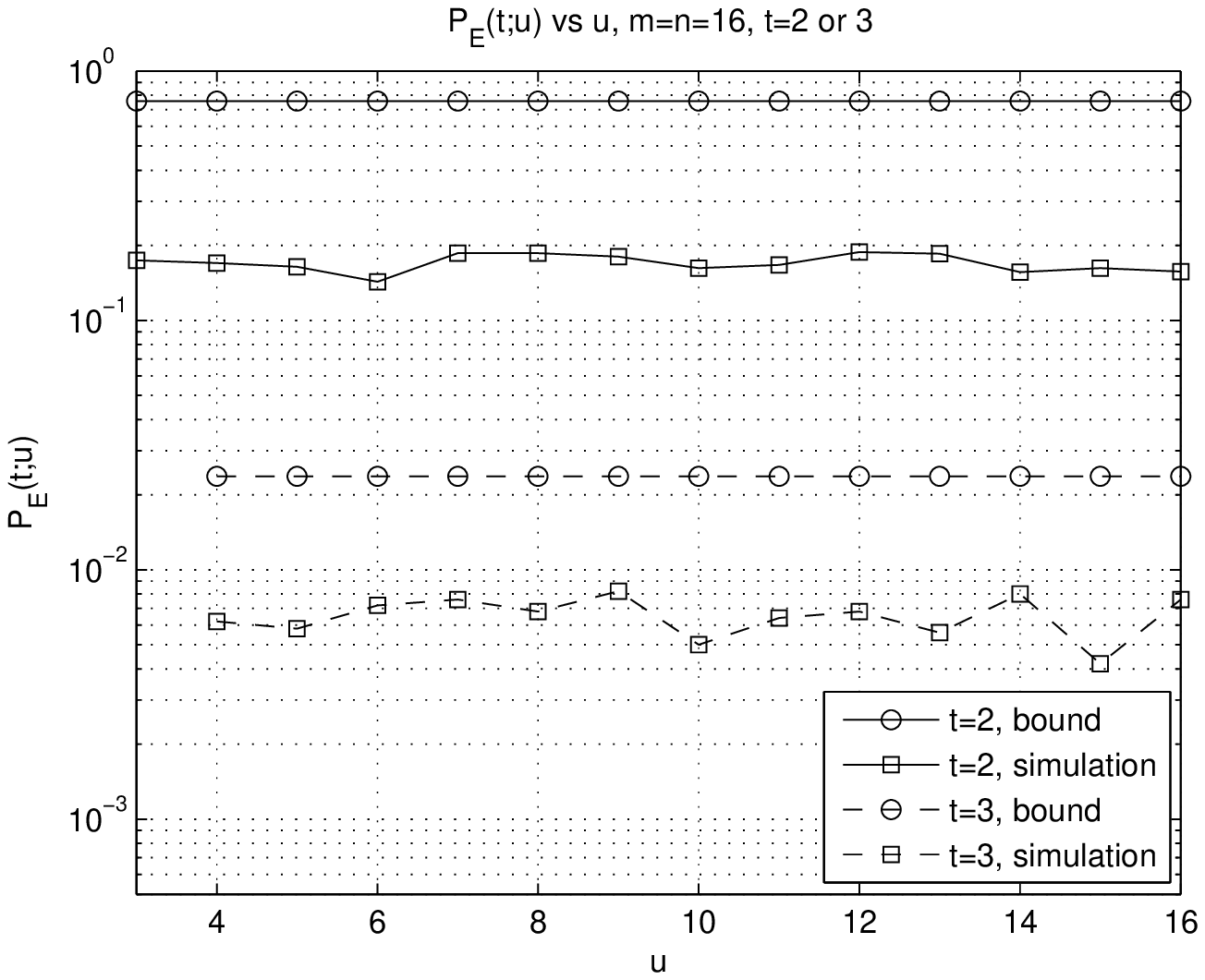}
\end{center}
\caption{Decoder error probability of an MRD code as a function of
$u$, with $q=2$, $m=n=16$, and $t$ equal to 2 or
3.}\label{fig:PE_MRDu}
\end{figure}

Note that our bound given in Proposition~\ref{prop:bound_PE2} does
not depend on $u$, and decreases exponentially with $t^2$. In
Figure~\ref{fig:PE_MRDt}, the decoder error probability is viewed as
a function of $t$ as $t$ varies from $1$ to $4$ and $u$ is set to
$n=16$. Note that when $t=1$, the bound in
Proposition~\ref{prop:bound_PE2} is trivial. We observe that both
the bound and the simulated decoder error probability decrease
exponentially with $t^2$. In Figure~\ref{fig:PE_MRDu}, the decoder
error probability is viewed as a function of $u$ as $t$ is set to
either $2$ or $3$ and $u$ varies from $t+1$ to $n=16$. Clearly, the
decoder error probability varies with $u$ somewhat, but the bound in
Proposition~\ref{prop:bound_PE2} is applicable to all values of $u$.

\bibliographystyle{IEEEtran}
\bibliography{gpt}

\end{document}